\documentclass[preprint,floatfix,superscriptaddress,a4paper,showpacs,showkeys,nofootinbib,notitlepage]{revtex4-2}

\usepackage{graphicx}
\usepackage[dvipsnames]{xcolor}
\usepackage[colorlinks=true,linktocpage=true,linkcolor=blue,citecolor=blue,allcolors=blue]{hyperref}
\usepackage[utf8]{inputenc}
\usepackage{indentfirst}
\usepackage{mathrsfs}
\usepackage{amsmath}
\usepackage{amssymb}
\usepackage[english]{babel}
\usepackage{url}
\usepackage[section]{placeins}

\newcommand{\edit}[1]{\ifmmode\textcolor{blue}{\boldsymbol{#1}}\else\textcolor{blue}{\textbf{#1}}\fi}

\begin{document}

\title{Strangeness-to-entropy ratio across the QCD crossover}

\author{Artemiy Lysenko}
\affiliation{Physics Department, Taras Shevchenko National University of Kyiv, 03022 Kyiv, Ukraine}
\affiliation{Bogolyubov Institute for Theoretical Physics, 03680 Kyiv, Ukraine}

\author{Mark I. Gorenstein}
\affiliation{Bogolyubov Institute for Theoretical Physics, 03680 Kyiv, Ukraine}

\author{Marek Gazdzicki}
\affiliation{Jan Kochanowski University, 25-369 Kielce, Poland}

\author{Roman Poberezhniuk}
\affiliation{Physics Department, University of Houston, Box 351550, Houston, Texas 77204, USA}
\affiliation{Bogolyubov Institute for Theoretical Physics, 03680 Kyiv, Ukraine}

\author{Volodymyr Vovchenko}
\affiliation{Physics Department, University of Houston, Box 351550, Houston, Texas 77204, USA}

\begin{abstract}
We study the absolute open-strangeness density $n_{|S|}$ of strongly interacting matter at vanishing conserved-charge chemical potentials. Unlike the net-strangeness susceptibility $\chi_2^S$, this quantity is not a derivative of the QCD partition function and requires a microscopic specification of the strangeness carriers. We use an equation of state that interpolates between an interacting hadron resonance gas and a quark-gluon plasma (QGP), with parameters fixed previously by lattice-QCD bulk thermodynamics. The ratio of $n_{|S|}$ to the entropy density $s$ follows the hadron-gas behavior below $T\simeq120$~MeV, rises above the ideal massless $2+1$-flavor limit of approximately $0.053$ across the crossover, and approaches this limit from above at high temperature. The same equation of state, minimally extended to finite $\mu_S$, captures the temperature dependence of the lattice-QCD $\chi_2^S$. For uncorrelated singly-strange Boltzmann particles, the strangeness fluctuation factor $F_S=\chi_2^S/(n_{|S|}/T^3)$ equals unity. Within the constructed equation of state, it deviates from unity by less than 10\%. This allows us to estimate $n_{|S|}/s$ from lattice-QCD data on $\chi_2^S$ and $s$. The lattice-based estimate reaches a maximum of $n_{|S|}/s\simeq0.063$ at $T\simeq220$~MeV, followed by a 20\% decrease towards the high-temperature limit.

\end{abstract}

\maketitle

\newpage

\section{Introduction}\label{intro}

Absolute strangeness, the sum over all strange and antistrange charges,  has long played an important role in experimental searches for the quark-gluon plasma (QGP). Enhanced absolute strangeness per nucleon created in nucleus-nucleus collisions, compared with proton-proton collisions, was proposed as a signature of QGP~\cite{Rafelski:1982pu,Koch:1986ud}. The underlying argument is that strange quarks can be produced efficiently through partonic reactions and may approach chemical equilibrium within the lifetime of the deconfined medium. 

A non-monotonic collision-energy dependence of the strangeness-to-pion ratio, known as the ``horn'', was predicted as a signal of the onset of deconfinement in Refs.~\cite{Gazdzicki:1998vd,Gazdzicki:2010iv}. It was subsequently observed in central Pb+Pb collisions, most prominently in the $\langle K^+\rangle/\langle\pi^+\rangle$ ratio~\cite{NA49:2002pzu,NA49:2007stj}.

Experimentally, the relative production of absolute strangeness is commonly characterized by ratios of strange-hadron multiplicities to pion multiplicities~\cite{Gazdzicki:1996pk,NA49:2007stj,STAR:2017sal,ALICE:2016fzo}. The kaon and hyperon yields serve as proxies for the abundance of strange and antistrange constituents, whereas the mean pion multiplicity is used as a proxy for the entropy produced in the collision~\cite{Gazdzicki:1995ze,Gazdzicki:2010iv}. The experimentally measured absolute strangeness-to-pion ratio may therefore be related, up to model-dependent corrections, to the theoretical absolute strangeness-to-entropy ratio~\cite{Gazdzicki:1995ze}. This relation assumes that the strangeness-to-entropy ratio is established early in the collision and is approximately conserved during the subsequent expansion~\cite{Gazdzicki:1998vd,Gazdzicki:2010iv}.

Unlike the entropy density, the absolute strangeness density is not defined by a derivative of the QCD partition function~\cite{Laine:2016hma,Borsanyi:2011sw}. Evaluating it requires a microscopic specification of the degrees of freedom assigned to the medium, so the density cannot be obtained uniquely from the bulk thermodynamics computed in lattice QCD. Lattice-QCD correlations of net strangeness with baryon number and electric charge indicate that strangeness is carried by hadrons up to the crossover and by weakly interacting quarks only above about twice the crossover temperature~\cite{Bazavov:2013dta}. Microscopic models that interpolate between hadronic and partonic degrees of freedom and are fitted to lattice-QCD results~\cite{Albright:2014gva,Lysenko:2026ctu} specify the carriers at all temperatures. In the following, we use the model we recently proposed~\cite{Lysenko:2026ctu} and calculate the ratio of absolute strangeness to entropy as a function of temperature at zero chemical potentials. We also calculate the net-strangeness susceptibility, which is known from lattice QCD, and use its relation to the absolute strangeness density to estimate the strangeness-to-entropy ratio from lattice-QCD data.

This paper is organized as follows. Section~\ref{sec:eos} summarizes the microscopic equation of state used throughout the study. Section~\ref{sec:net-strangeness-test} evaluates the net-strangeness susceptibility and compares it with lattice-QCD results. Section~\ref{sec:absolute-ratio} presents the temperature dependence of the absolute open-strangeness-to-entropy ratio. Section~\ref{sec:absolute-versus-fluctuations} examines the relation between absolute open strangeness and net-strangeness fluctuations. Finally, Sec.~\ref{sec:summary} summarizes the main results and outlines their connection to experimental studies.

\section{Equation of state}
\label{sec:eos}

All calculations presented below are performed at vanishing conserved-charge chemical potentials, $\mu_B=\mu_Q=\mu_S=0$, using the microscopic equation of state (EoS) constructed and fitted to lattice-QCD thermodynamics in Ref.~\cite{Lysenko:2026ctu}.  Here we summarize only those elements of the EoS that are required for the calculation of strangeness observables.

The pressure is obtained by smoothly interpolating between a hadron resonance gas (HRG) contribution, $p_H$, and a quark-gluon plasma (QGP) contribution, $p_Q$,
\begin{equation}
    p(T)=\mathscr{K}(T)p_Q(T)
    +\left[1-\mathscr{K}(T)\right]p_H(T),
    \label{EoS}
\end{equation}
with the switching function
\begin{equation}
    \mathscr{K}(T)
    =\exp\!\left[-\left(\frac{T_0}{T}\right)^k\right].
    \label{switching-function}
\end{equation}
Thus, $\mathscr{K}(T)$ and $1-\mathscr{K}(T)$ are the temperature-dependent weights of the QGP and HRG pressures, respectively. The interpolation procedure~(\ref{EoS}) and the analytical form of the switching function~(\ref{switching-function}) were suggested in Ref.~\cite{Albright:2014gva}.

The hadronic component is described by the quantum van der Waals HRG model (QvdW HRG)~\cite{Vovchenko:2016rkn}, which supplements the ideal hadron resonance gas~\cite{Dashen:1969ep} with attractive mean-field and excluded-volume~\cite{Rischke:1991ke} interactions. Van der Waals interactions are included among baryons and, separately, among antibaryons, with the attraction and excluded-volume parameters $a=329$~MeV~fm$^3$ and $b=3.42$~fm$^3$ fixed by the ground-state properties of nuclear matter~\cite{Vovchenko:2015vxa,Vovchenko:2015pya,Vovchenko:2016rkn}.  The latter value corresponds to the (anti)baryon hard-core radius $r_B=0.59$~fm. The model is extended by excluded-volume repulsion between mesons, characterized by a common meson hard-core radius $r_M$~\cite{Lysenko:2026ctu}. The HRG calculations use the PDG2020 hadron list implemented in the Thermal-FIST package~\cite{Vovchenko:2019pjl}. The meson repulsion has little effect on the HRG pressure below $T\simeq160$~MeV but becomes important at higher temperatures~\cite{Lysenko:2026ctu}.

The QGP component consists of an ideal gas of gluons and $2+1$ flavors of quarks supplemented by a phenomenological interaction term,
\begin{equation}
    p_Q(T)=p_Q^{\rm id}(T)-AT^3.
    \label{QGP}
\end{equation}
The up- and down-quark masses are neglected, while the strange-quark mass is fixed at $m_s=93.5$~MeV~\cite{ParticleDataGroup:2024cfk}.  The correction $-AT^3$ accounts phenomenologically for the interaction effects which vanish relative to the leading $T^4$ contribution at asymptotically high temperature.

For integer values of the exponent $k$, the three parameters $r_M$, $A$, and $T_0$ were determined in Ref.~\cite{Lysenko:2026ctu} from a fit to the lattice-QCD trace anomaly in the temperature interval $100\leq T\leq510$~MeV. The best fits are obtained for $k=4$ and $k=5$:
\begin{equation} \label{eos-parameters}
\begin{split}
    k=4, \qquad T_0=179 \ \text{MeV}, \qquad A=587 \ \text{MeV}, \qquad r_M=0.46 \ \text{fm}, \\
    k=5, \qquad T_0=189 \ \text{MeV}, \qquad A=624 \ \text{MeV}, \qquad r_M=0.34 \ \text{fm}.
\end{split}
\end{equation}
The QGP and HRG weights in EoS (\ref{EoS}) are equal, $\mathscr{K}=1/2$, at $T=196$~MeV for $k=4$ and at $T=203$~MeV for $k=5$. Since the two sets of values describe the lattice data for the trace anomaly equally well (see Ref.~\cite{Lysenko:2026ctu}), in the next section we choose a specific set from the condition of best describing the susceptibility of the net-strangeness $\chi_2^S$.

\section{Net-strangeness susceptibility}
\label{sec:net-strangeness-test}

The parameters of the EoS introduced in Sec.~\ref{sec:eos} were determined in Ref.~\cite{Lysenko:2026ctu} exclusively from thermodynamic quantities at vanishing conserved-charge chemical potentials. Consequently, the $\mu_S$ dependence is not fixed by the original EoS~(\ref{EoS}), and neither is the second-order net-strangeness susceptibility,
\begin{equation}
    \chi_2^S
    =\left.
      \frac{\partial^2(p/T^4)}{\partial(\mu_S/T)^2}
      \right|_{\mu_B=\mu_Q=\mu_S=0}
    =\frac{\langle N_S^2\rangle}{VT^3},
    \label{chi-2}
\end{equation}
where $N_S$ is the net strangeness in the volume $V$ and $\langle N_S\rangle=0$ at vanishing chemical potentials. We introduce the minimal finite-$\mu_S$ extension, $p_Q(T)\rightarrow p_Q(T,\mu_S)$ and $p_H(T)\rightarrow p_H(T,\mu_S)$ at $\mu_B=\mu_Q=0$, neglecting the $\mu_S$ dependence of the switching function $\mathscr{K}$ and of the phenomenological QGP interaction term $-AT^3$.

\begin{figure}[h!]
    \centering
    \includegraphics[width=.70\textwidth]{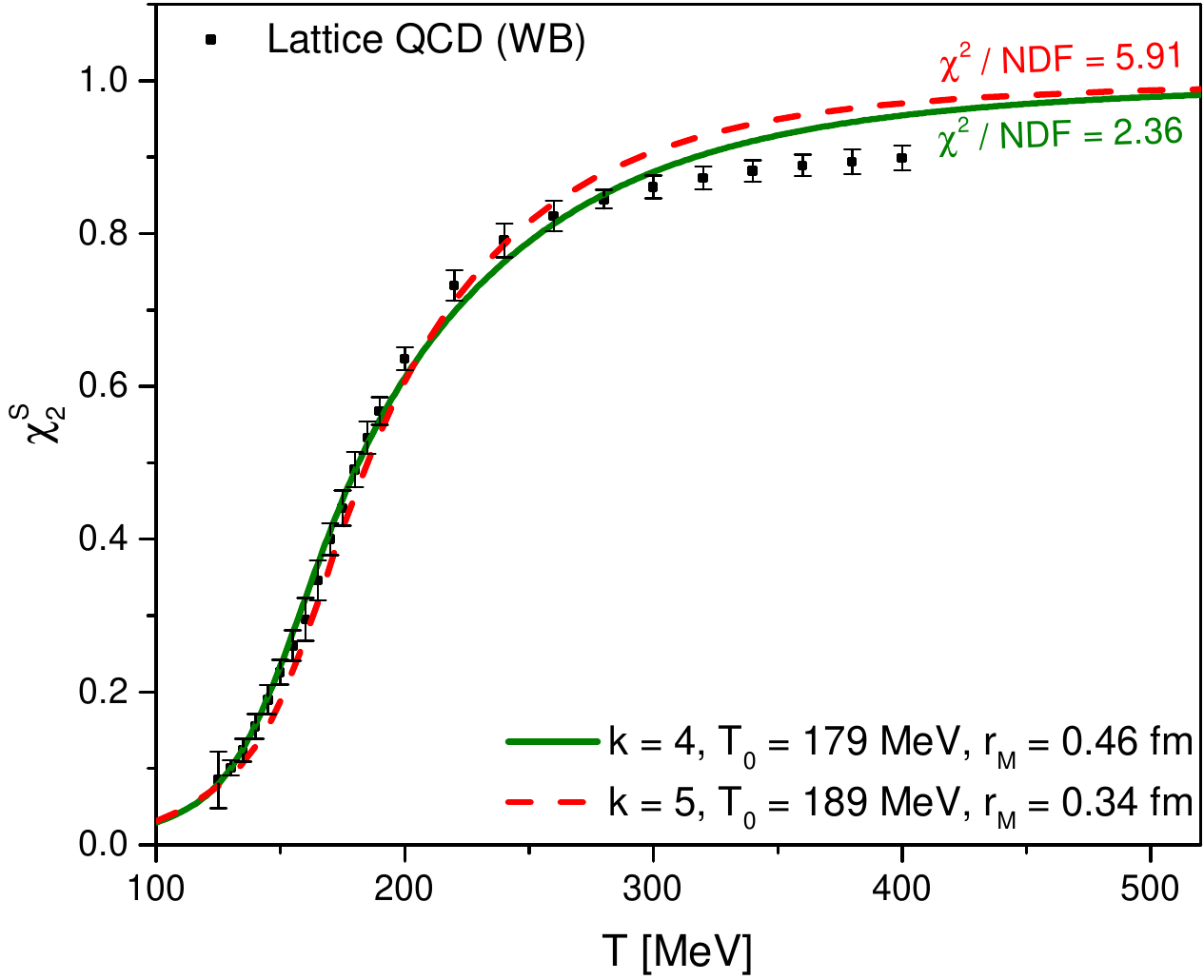}
    \caption{Temperature dependence of the net-strangeness susceptibility $\chi_2^S$. The green solid and red dashed curves show the full EoS results obtained using Eq.~(\ref{fullchi2S}) with the $k=4$ and $k=5$ parameter sets of Eq.~(\ref{eos-parameters}), respectively. The Wuppertal–Budapest lattice-QCD results~\cite{Borsanyi:2011sw} are shown by black symbols.}
    \label{figChi}
\end{figure}

Under these assumptions, differentiating Eq.~(\ref{EoS}) with respect to $\mu_S$ gives
\begin{equation}
    \chi_2^S(T)
    =\mathscr{K}(T)\chi_2^{S[Q]}(T)
     +\left[1-\mathscr{K}(T)\right]\chi_2^{S[H]}(T),
    \label{fullchi2S}
\end{equation}
where
\begin{equation}
    \chi_2^{S[X]}(T)
    =\left.
      \frac{\partial^2[p_X(T,\mu_S)/T^4]}
           {\partial(\mu_S/T)^2}
      \right|_{\mu_S=0},
    \qquad X\in\{H,Q\}.
    \label{component-chi2S}
\end{equation}
For the QGP component this yields
\begin{equation}
    \chi_2^{S[Q]}
    =\frac{d_s}{2\pi^2T^3}
      \int_0^\infty d\kappa \, \kappa^2 f_s(\kappa)\left[1-f_s(\kappa)\right],
    \qquad
    f_s(\kappa)
    =\left[
        \exp\!\left(\frac{\sqrt{\kappa^2+m_s^2}}{T}\right)+1
      \right]^{-1}.
    \label{chi2S-QGP}
\end{equation}
Here $d_s=12$ includes spin, color, and the strange-quark and antiquark degrees of freedom. The HRG contribution $\chi_2^{S[H]}$ is obtained similarly by differentiating the QvdW HRG pressure with respect to $\mu_S$.

Figure~\ref{figChi} compares the temperature dependence of the net-strangeness susceptibility obtained with two parameterizations of the full EoS~(\ref{eos-parameters}) to the Wuppertal–Budapest lattice-QCD results. Both parameterizations reproduce the overall rapid increase of $\chi_2^S$ through the crossover region and its subsequent approach towards a high-temperature plateau. However, the model results increasingly overestimate the lattice data at higher temperatures, with the discrepancy reaching about 7\% at $T = 400$ MeV. The $k=4$ parameterization provides a better description of the lattice results, with $\chi^2/\mathrm{NDF}=2.36$, compared to $\chi^2/\mathrm{NDF}=5.91$ for the $k=5$ parameterization. Here NDF $=25$ is the number of lattice points, and their uncertainties are treated as uncorrelated.

In the following analysis, we use the $k=4$ parameterization, which better describes the lattice-QCD susceptibility data.

\section{Absolute open-strangeness-to-entropy ratio}
\label{sec:absolute-ratio}

For the HRG component, the \emph{absolute open-strangeness density} is defined by weighting the primordial equilibrium density of each hadron species by the magnitude of its net strangeness charge,
\begin{equation}
    n_{|S|}^{H}(T)
    =\sum_i \lvert q_i^S\rvert\,n_i(T),
    \label{S-HRG_2}
\end{equation}
where $q_i^S$ and $n_i$ are the net strangeness charge and the number density of species $i$, respectively. The sum includes stable hadrons and resonances before their decays. Hidden-strangeness particles with $q_i^S=0$, such as the $\phi(s\bar{s})$, do not contribute, although they contain strange and antistrange valence quarks. Thus, $n_{|S|}$ is smaller than the total density of strange and antistrange valence quarks.

\begin{figure}[h!]
    \centering
    \includegraphics[width=.70\textwidth]{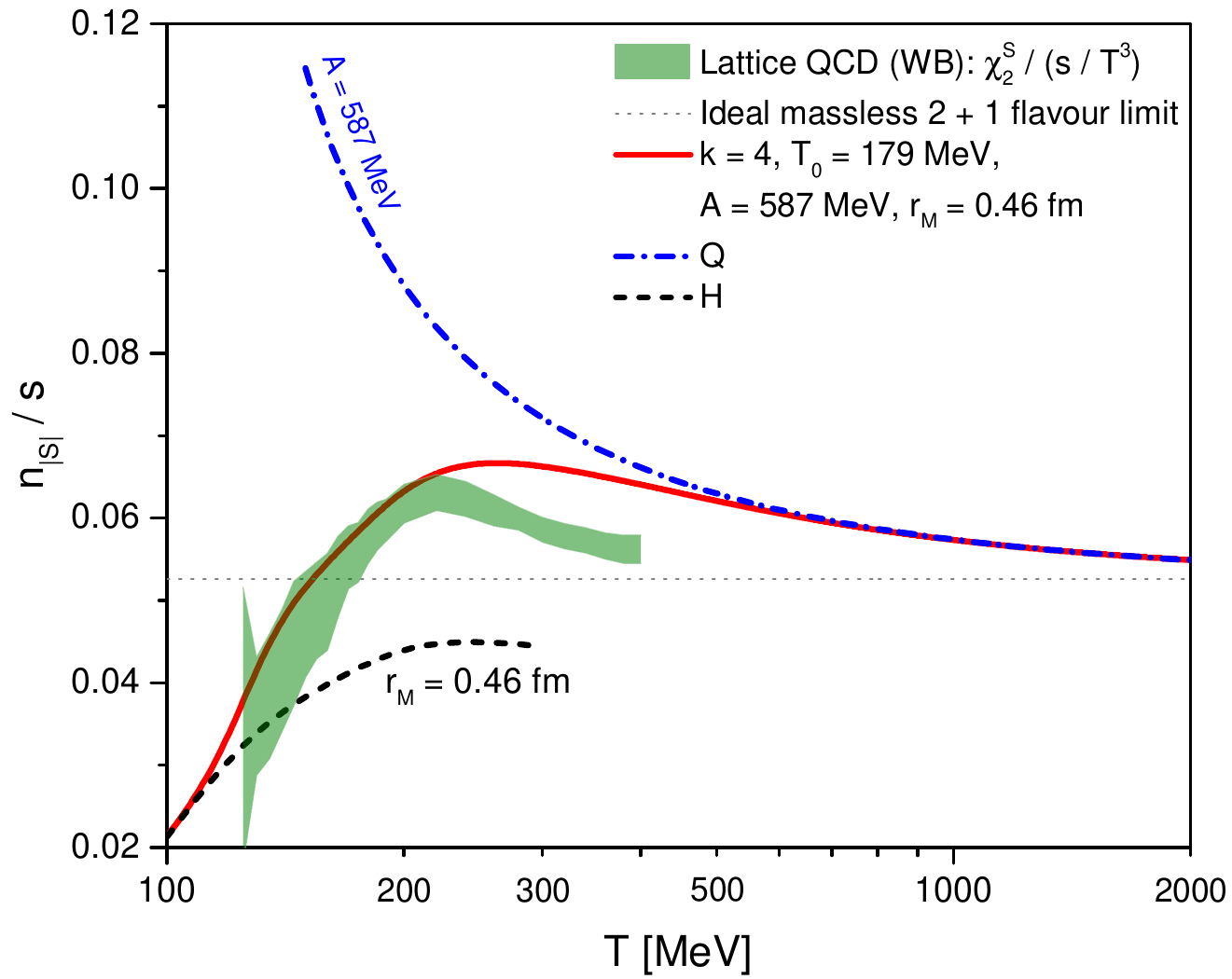}
    \caption{Temperature dependence of the absolute open-strangeness-to-entropy ratio $n_{|S|}/s$ obtained from Eqs.~(\ref{StoS}) and~(\ref{entropy-ratio}) (red solid line). The HRG ratio $n_{|S|}^{H}/s_H$ and the QGP ratio $n_{|S|}^{Q}/s_Q$ are shown by the black dashed and blue dash-dotted lines, respectively. The dotted line is the ideal massless limit of Eq.~(\ref{asymptotic-ratio}). The green band shows the estimate of $n_{|S|}/s$ obtained by setting $F_S=1$ in Eq.~(\ref{strangeness-fluctuation-factor}) and using lattice-QCD data for $\chi^S_2$~\cite{Borsanyi:2011sw} and $s/T^3$~\cite{Borsanyi:2013bia}. The green band includes the lattice uncertainties.}
    \label{figStr}
\end{figure}

For the QGP component, strange quarks and antiquarks are counted directly,
\begin{align}
    n_{|S|}^{Q}(T)
    &=n_s+n_{\bar{s}}
      =\frac{d_s}{2\pi^2}
        \int_0^\infty d\kappa\,\kappa^2 f_s(\kappa).
    \label{S-QGP}
\end{align}
The full absolute open-strangeness density is defined by the same interpolation weights as the pressure,
\begin{equation}
    n_{|S|}(T)
    =\mathscr{K}(T)n_{|S|}^{Q}(T)
     +\left[1-\mathscr{K}(T)\right]n_{|S|}^{H}(T).
    \label{StoS}
\end{equation}
The same weights enter $\chi_2^S$ in Eq.~(\ref{fullchi2S}), so that both quantities are composed of their hadronic and partonic parts in the same way.

The entropy density is obtained by differentiating the full pressure~(\ref{EoS}),
\begin{align}
    s(T)
    =\frac{dp}{dT} 
    =\mathscr{K}(T)s_Q(T)
      +\left[1-\mathscr{K}(T)\right]s_H(T)
      +\mathscr{K}'(T)\left[p_Q(T)-p_H(T)\right],
    \label{entropy-ratio}
\end{align}
where $s_Q=dp_Q/dT$, $s_H=dp_H/dT$, and
\begin{equation}
    \mathscr{K}'(T)
    =\frac{k}{T}\left(\frac{T_0}{T}\right)^k\mathscr{K}(T).
\end{equation}
The last term in Eq.~(\ref{entropy-ratio}) is the crossing contribution caused by the temperature dependence of the interpolation weights.

When $T\to\infty$, $m_s/T \to 0$, while the interaction term becomes subleading, $-AT^3/T^4=-A/T\to0$.  For an ideal massless gas of gluons and $2+1$ quark flavors,
\begin{equation}
    \frac{n_{|S|}^{Q}}{T^3}
      \longrightarrow \frac{9\,\zeta(3)}{\pi^2},
    \qquad
    \frac{s_Q}{T^3}
      \longrightarrow \frac{19\pi^2}{9},
\end{equation}
and hence
\begin{equation}
    \lim_{T\to\infty}\frac{n_{|S|}}{s}
    =\frac{81\,\zeta(3)}{19\pi^4}
    \simeq0.053.
    \label{asymptotic-ratio}
\end{equation}

The temperature dependence of the absolute open-strangeness-to-entropy ratio $n_{|S|}(T)/s(T)$ is shown in Fig.~\ref{figStr}. At low temperatures, $T \lesssim 120$ MeV, the full EoS result closely follows the HRG contribution. As temperature increases, the QGP weight becomes progressively more important, leading to a rapid increase in the full ratio through the crossover region. The ratio reaches a maximum of approximately $0.067$ at $T \simeq 260$ MeV. At higher temperatures, it gradually decreases to $\approx 0.062$ at $T=500$ MeV and approaches the asymptotic high-temperature limit of $\approx 0.053$. The absolute open-strangeness-to-entropy ratio for HRG and QGP is plotted separately in Fig.~\ref{figStr} by the black dashed and blue dash-dotted lines, respectively. The HRG ratio exhibits a maximum of $0.045$ at $T \simeq 250$ MeV. The QGP ratio lies above the HRG ratio at all temperatures and gradually approaches the ideal massless $2+1$ flavor limit as the strange-quark mass becomes negligible compared with the temperature. A QGP ratio above the hadron-gas ratio was obtained earlier in Ref.~\cite{Kuznetsova:2006bh}, where lowest-order perturbative corrections reduce the entropy density more strongly than the strange-quark density. For temperatures between $140$ and $260$~MeV, that calculation gives about $0.032$ strange-quark pairs per unit entropy, that is, $n_{|S|}/s\approx0.064$. The green band in Fig.~\ref{figStr} is an estimate based on lattice-QCD data, which is discussed in Sec.~\ref{sec:absolute-versus-fluctuations}.

\section{Relation between absolute open strangeness and \\ net-strangeness fluctuations}
\label{sec:absolute-versus-fluctuations}

The absolute open-strangeness density $n_{|S|}$ and the net-strangeness susceptibility $\chi_2^S$ quantify different properties of the system.  The former measures the mean abundance of open-strangeness carriers, whereas the latter measures fluctuations of the conserved net-strangeness charge.  To compare them, we introduce the dimensionless strangeness fluctuation factor
\begin{equation}
    F_S(T)
    \equiv
    \frac{\chi_2^S(T)}{n_{|S|}(T)/T^3}.
    \label{strangeness-fluctuation-factor}
\end{equation}

Consider an ideal Boltzmann gas of mutually independent particle species, all carrying $\lvert q_i^S\rvert=1$. In the grand-canonical ensemble, the multiplicity of every species is then Poisson distributed. Consequently, the variance of the net-strangeness charge equals the mean number of strange plus antistrange carriers:
\begin{equation}
    \chi_2^{S[P]}
    =\frac{n_{|S|}^{P}}{T^3},
    \qquad
    F_S^{P}=1.
    \label{Poisson}
\end{equation}
Multistrange particles, quantum statistics, and interactions produce deviations from this baseline.

Let $n_1$, $n_2$, and $n_3$ denote the total particle-plus-antiparticle densities of hadrons carrying $\lvert q_i^S\rvert=1$, 2, and 3, respectively. For an ideal Boltzmann mixture with independent Poisson multiplicities~\cite{Karsch:2010ck},
\begin{equation}
    \frac{n_{|S|}^{H}}{T^3}
    =\frac{n_1+2n_2+3n_3}{T^3},
    \qquad
    \chi_2^{S[H]}
    =\frac{n_1+4n_2+9n_3}{T^3}.
    \label{multistrange}
\end{equation}
Thus, the absolute density weights a species linearly with $\lvert q_i^S\rvert$, whereas the susceptibility weights it quadratically. Multistrange states therefore enhance $\chi_2^S$ more strongly than $n_{|S|}/T^3$ and yield $F_S>1$ even in an otherwise ideal Boltzmann gas.

Quantum statistics and interactions introduce additional corrections. Bose statistics enhances strange-meson fluctuations, and attractive van der Waals interactions enhance baryonic fluctuations; both effects tend to increase $F_S$. Fermi statistics suppresses strange-baryon fluctuations, whereas excluded-volume repulsion suppresses fluctuations of both mesonic and (anti)baryonic components; these effects tend to decrease $F_S$. Near $T=150$~MeV, the result for hadronic matter is
\begin{equation}
    \chi_2^{S[H]}
    \simeq1.10\,\frac{n_{|S|}^{H}}{T^3},
    \label{FS-hadronic}
\end{equation}
with the enhancement generated almost entirely by multistrange baryons. The combined contribution of quantum statistics and interactions remains below the percent level at this temperature. At higher temperature, excluded-volume effects in the HRG component grow rapidly and suppress fluctuations relative to the absolute open-strangeness density.

\begin{figure}[h!]
    \centering
    \includegraphics[width=.70\textwidth]{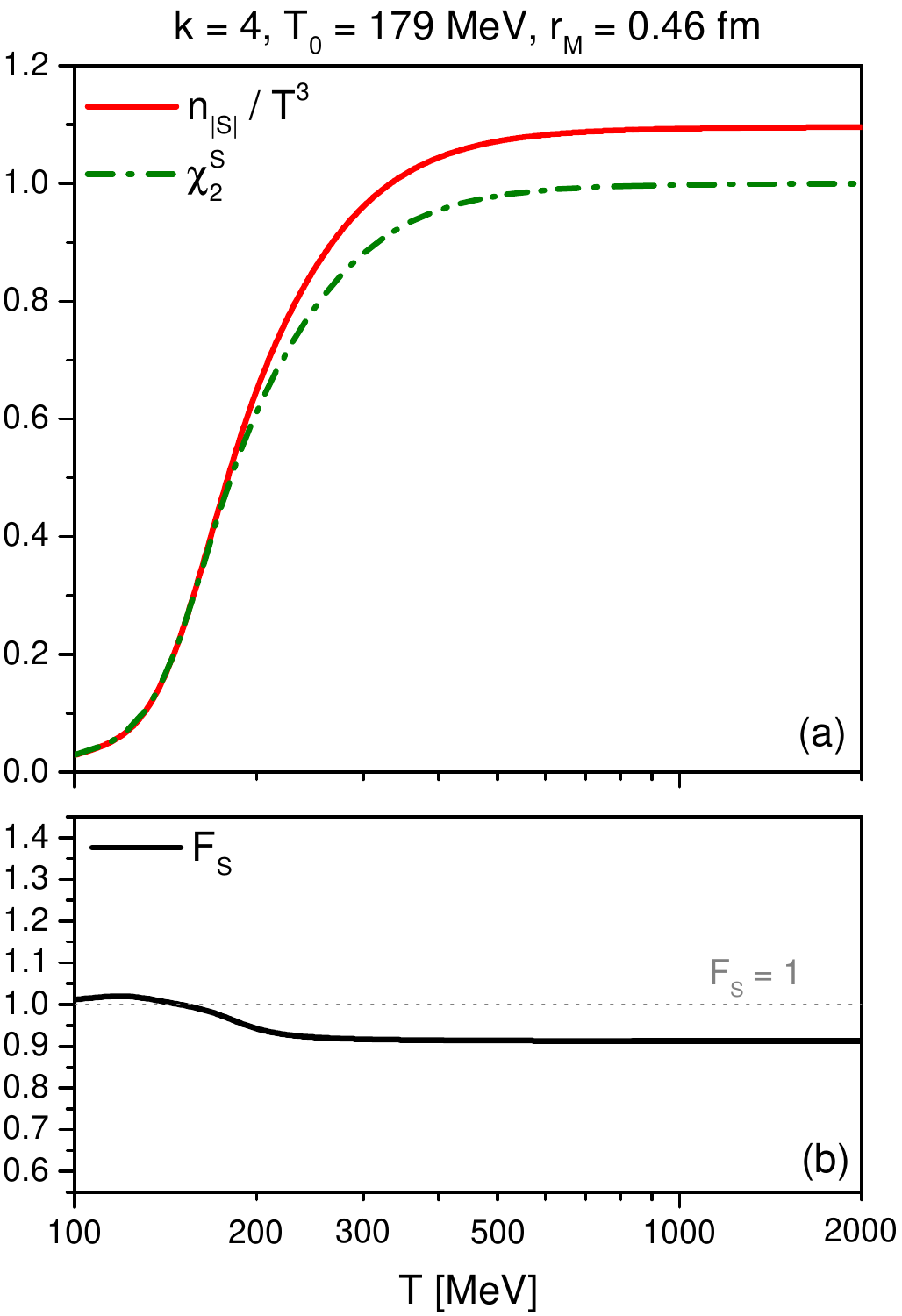}
    \caption{Model results for the relation between absolute open strangeness and net-strangeness fluctuations. Upper panel (a): temperature dependence of $n_{|S|}/T^3$ from Eq.~(\ref{StoS}) and $\chi_2^S$ from Eq.~(\ref{fullchi2S}). Lower panel (b): the strangeness fluctuation factor $F_S=\chi_2^S/[n_{|S|}/T^3]$. The horizontal line at $F_S=1$ marks the independent, singly-strange Boltzmann baseline of Eq.~(\ref{Poisson}).}
    \label{figSchi}
\end{figure}

For the QGP component, the absolute density in Eq.~(\ref{S-QGP}) contains the Fermi--Dirac occupation factor $f_s(\kappa)$, whereas the fluctuation of each mode given by Eq.~(\ref{chi2S-QGP}) also contains the Pauli-blocking factor, $1-f_s(\kappa)$. Since $0<f_s(\kappa)<1$, one has $f_s(\kappa)[1-f_s(\kappa)]<f_s(\kappa)$ mode by mode.  Fermi statistics therefore suppresses net-strangeness fluctuations more strongly than the mean strange-plus-antistrange abundance. At $T\simeq400$~MeV,
\begin{equation}
    \chi_2^{S[Q]}
    \simeq0.91\,\frac{n_{|S|}^{Q}}{T^3}.
    \label{partonic-suppression}
\end{equation}
For massless strange quarks, the limiting fluctuation factor is
\begin{equation}
    \lim_{T\to\infty}F_S^{Q}(T)
    =\frac{\pi^2}{9\zeta(3)}
    \simeq0.912.
    \label{asymptotic-fluctuation-factor}
\end{equation}
The deviation from the Poisson baseline therefore persists even in an ideal ultrarelativistic QGP.

Figure~\ref{figSchi}(a) shows the temperature dependence of the absolute open-strangeness density, normalized by $T^3$, and the net-strangeness susceptibility $\chi_2^S$. The two quantities are nearly equal in the dilute hadronic regime, consistent with the Poisson baseline for independent singly-strange Boltzmann particles (see Eq.~(\ref{Poisson})). As the temperature increases through the crossover region, their difference becomes increasingly pronounced: $n_{|S|}/T^3$ exceeds $\chi_2^S$ by about 10\%, and this persists at high temperatures.

Figure~\ref{figSchi}(b) shows the corresponding temperature dependence of the strangeness fluctuation factor $F_S$. At low temperatures, $F_S$ remains close to unity and exhibits only a small enhancement, reaching a maximum of approximately $1.02$ at $T \simeq 120$ MeV. With increasing temperature, the weight of the QGP component grows, and $F_S$ decreases and crosses unity at $T \simeq 150$ MeV. It approaches the ideal massless-quark limit of $0.912$ at high temperatures, becoming close to this value already around $T=500$ MeV. The deviation of $F_S$ from unity remains below 10\% over the entire temperature range considered.

This fact allows us to estimate $n_{|S|}/s$ from lattice QCD. Setting $F_S=1$, we replace $n_{|S|}/T^3$ by the lattice-QCD susceptibility $\chi_2^S$~\cite{Borsanyi:2011sw} and divide it by the lattice-QCD entropy density $s/T^3$~\cite{Borsanyi:2013bia}. The result is the lattice estimate of the absolute open-strangeness-to-entropy ratio, shown in Fig.~\ref{figStr} as the green band.

The maximum of the lattice estimate is reached at $T \simeq 220$ MeV and is approximately $0.063$, which is slightly less than the model result (red solid line in Fig.~\ref{figStr}, $0.067$ at $T \simeq 260$ MeV). At the same time, in the temperature range $T\approx130-210$ MeV, the model result agrees with the lattice estimate (the red solid line lies within the green band in Fig.~\ref{figStr}).

\section{Summary}
\label{sec:summary}

We have studied two complementary equilibrium observables of strangeness in strongly interacting matter: fluctuations of the conserved net-strangeness charge and the mean abundance of open-strangeness carriers. We performed the calculations at vanishing conserved-charge chemical potentials using the EoS of Ref.~\cite{Lysenko:2026ctu}, which interpolates between an interacting HRG and an interacting QGP. 

To calculate the net-strangeness susceptibility $\chi_2^S$, we introduced the $\mu_S$ dependence into the equation of state of Ref.~\cite{Lysenko:2026ctu}. Without retuning the model parameters, the extended EoS reproduces the overall temperature dependence of the susceptibility, as demonstrated in Fig.~\ref{figChi}. A comparison of the two parameterizations shows that the $k=4$ parameter set better describes the Wuppertal–Budapest lattice-QCD results~\cite{Borsanyi:2011sw} than the $k=5$ parameter set, although deviations from the lattice data remain, particularly at high temperatures.

We calculate the temperature dependence of the absolute open-strangeness-to-entropy ratio $n_{|S|}/s$, shown in Fig.~\ref{figStr}. At low temperatures ($\lesssim 120$ MeV), the ratio increases with temperature, closely following the HRG contribution. It reaches a maximum of approximately $0.067$ at $T \simeq 260$ MeV, then gradually decreases toward the asymptotic high-temperature limit of $0.053$. This limit corresponds to an ideal massless gas of gluons and $2+1$ quark flavors.

For singly-strange Boltzmann particles with Poisson multiplicity distributions, $n_{|S|}/T^3$ equals $\chi_2^S$ (see Eq.~(\ref{Poisson})). We quantify the difference between the two quantities by the strangeness fluctuation factor $F_S=\chi_2^S/(n_{|S|}/T^3)$, shown in Fig.~\ref{figSchi}. At low temperatures, the hadronic gas dominates, and $\chi_2^S$ is larger than $n_{|S|}/T^3$ mainly due to multistrange hadrons, leading to a maximum $F_S \simeq 1.02$ at $T \approx 120$~MeV. With increasing temperature, the contribution of the QGP and the influence of Fermi statistics of strange quarks increase. Therefore, $F_S$ crosses unity at $T\simeq 150$~MeV and approaches the ideal limit of $0.912$ for massless quarks. Despite the deviation from the Poisson baseline, $F_S$ remains within approximately 10\% of unity (see Fig.~\ref{figSchi}(b)). This allows us to estimate $n_{|S|}/s$ from the lattice-QCD data for $\chi_2^S$ and $s/T^3$. The lattice-based estimate, shown in Fig.~\ref{figStr} as the green area, peaks at $n_{|S|}/s \simeq 0.063$ at $T \simeq 220$ MeV. In the temperature range from $130$ to $210$~MeV, the model calculation (red solid line in Fig.~\ref{figStr}) coincides with the lattice estimate.

The connection of the results presented here to experimental data remains necessarily indirect. The present calculation describes primordial equilibrium densities in a grand-canonical system, whereas experiments measure final-state hadrons within finite acceptance.  Resonance decays and feed-down, efficiencies, global conservation laws~\cite{Vovchenko:2020tsr,Vovchenko:2020gne}, and neutral or unmeasured particles must be incorporated before a quantitative comparison with strange-hadron multiplicities and their cumulants can be made. Extending the microscopic EoS consistently to finite baryon density and modeling hadronization and the subsequent hadronic stage are therefore the principal next steps toward confronting the predicted $n_{|S|}/s$ and $F_S$ with the data.

\begin{acknowledgments}
We thank M. Adzhymambetov and V. Kuznietsov for fruitful discussions. M.I.G. and A.L. acknowledge support from the National Research Foundation of Ukraine under grant No. 2025.07/0050.
\end{acknowledgments}

\bibliography{bibliography}

\end{document}